\documentclass[fleqn,12pt,twoside]{article}
\usepackage{espcrc1}

\usepackage{graphicx}
\usepackage{epsfig}

\usepackage[figuresright]{rotating}

\def\h#1{{\hspace*{#1}}}
\def\v#1{{\vspace*{#1}}}

\title{Structure Functions and Parton Distributions}

\author{Jianwei Qiu\address{Department of Physics and Astronomy,
                            Iowa State University,\\
                            Ames, Iowa 50011, U.S.A.}\thanks{{\tt 
        jwq@iastate.edu.}  I thank Eskola and Salgado for useful
                            discussions and figures.  This work is
                            supported in part by the United States 
                            Department of Energy under Grant 
                            No. DE-FG02-87ER40371.}}
\begin{document}

% typeset front matter
\maketitle

\begin{abstract}
In this talk, I review the status of theoretical understanding of 
nuclear structure functions and parton distributions
and discuss the constraints on nuclear parton distributions
from existing data and the global QCD analysis.
\end{abstract}

%%%%%%%%%%%%%%%%%%%%%%%%%%%%%%%%%%%%%%%%%%%%%%%%%%%%%%%%%%%%%%%%%%%%%%%%
\section{Introduction}

Much of the predictive content of perturbative QCD (pQCD) treatment
of the hadronic hard scattering is contained in factorization theorems 
\cite{Collins:gx}.  Their purpose is to separate long- from 
short-distance effects in scattering amplitudes.  They supply 
perturbatively uncalculable long-distance effects with physics
content in terms of well-defined matrix elements, which allows them 
to be measured experimentally or by numerical simulation.  They
also define the normalization of short-distance factors, which allows
them to be calculated perturbatively.  Predictions follow when
processes with different hard scatterings but the same nonperturbative
matrix elements are compared.  Thus, quark and gluon distributions
measured in deep inelastic scattering may be used to normalize
the Drell-Yan or jet cross section.

Deep inelastic lepton scattering has long been regarded as the cleanest
probe of constituent substructure.  Considerable interest, therefore, 
greeted the observation \cite{EMC83} by European Muon Collaboration (EMC)
that the structure function $F_2(x_B,Q^2)$ of an iron nucleus differs
in significant ways as a function of $x_B$ from that for deuterium. 
It was the discovery of the EMC effect that opened a door 
for systematic study of QCD dynamics in a nuclear environment, which
has led to many new QCD phenomena, e.g., shadowing, saturation, 
and color glass condensate.   
 
In this talk, I review our abilities and limitations to generalize 
the pQCD factorization theorems to the hard scattering involving nuclei. 
Nuclear parton distributions are heavily used in phenomenological 
description of hard processes in heavy ion reactions at SPS and RHIC 
energies and in calculating predictions at the LHC energies.  
According to the factorization theorems, nuclear dependence of parton 
distributions should be universal or process independent.  
I discuss the constraints on nuclear parton distributions 
from existing data and the global QCD analysis.  

%%%%%%%%%%%%%%%%%%%%%%%%%%%%%%%%%%%%%%%%%%%%%%%%%%%%%%%%%%%%%%%%%%%%%%%%
\section{Structure Functions and Parton Distributions}

%=======================================================================
\subsection{Structure functions}

An inclusive deep inelastic scattering (DIS) process is generically 
of the form, $\ell(E)+h(p) \Longrightarrow \ell(E')+X$, where $\ell$ 
represents a lepton, $h$ a hadron (a nucleon or nucleus), and $X$ an 
arbitrary hadronic final-state.  The process, illustrated in Fig.~1(a),
is initiated by the exchange of a virtual photon (or a vector boson in 
general).  In DIS, the momentum transfer between the lepton and the 
hadron, $q$, is spacelike, $-q^2 = Q^2$.  The Bjorken scaling variable is
defined as $x_B=\frac{Q^2}{2p\cdot q} = \frac{Q^2}{2m_h \nu}$, 
where $\nu$ is the energy transferred from the lepton to the hadron 
in the hadron (target) rest frame, $\nu={E-E'}$.  In the same frame,
$Q^2 = 4{EE'}\sin^2(\theta/2)$ with the lepton scattering angle 
$\theta$.

\v{-0.8cm}
\begin{figure}[ht]
\centerline{\includegraphics[width=15cm]{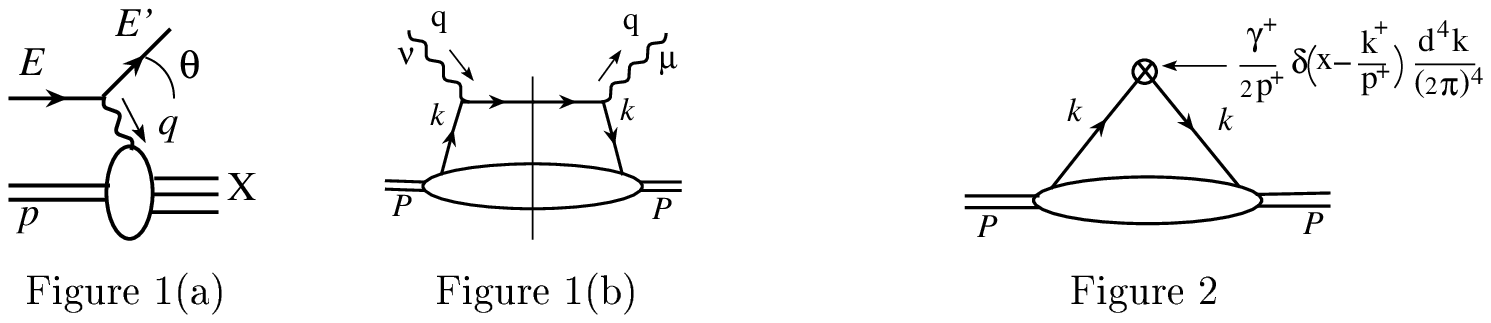}}
\v{-0.7cm}

\end{figure}

The DIS cross section with unpolarized beam and target can be written 
in the one photon exchange approximation as \cite{Guo:2001tz}
\begin{equation}
\frac{d\sigma^{\rm DIS}}{dx_B dQ^2}
= \frac{\sigma_{\rm Mott}}{EE'}
\frac{\pi {F_2(x_B,Q^2)}}{\epsilon x_B}
\left[\frac{1+\epsilon R(x_B,Q^2)}{1+R(x_B,Q^2)}\right]\ ,
\label{dis-x}
\end{equation}
where $\epsilon$ is a kinematic parameter and $R$ is the ratio of 
longitudinal to transverse photon-hadron cross section,
$R = \sigma_L/\sigma_T
=[\left(1+4x_B^2m^2/Q^2\right)F_2 - 2x_B F_1]/[2x_B F_1]$. 
The functions $F_1$ and $F_2$ are called {\it structure functions} 
and defined as
\begin{equation}
F_i(x_B,Q^2) \equiv e_i^{\mu\nu}\, 
W_{\mu\nu}(x_B,Q^2)
\label{f-wmn}
\end{equation}
with $i=1,2$ and projection tensors 
$e_i^{\mu\nu}$ given in Ref.~\cite{Guo:2001tz}.  
The $W_{\mu\nu}(x_B,Q^2)$ is 
the DIS hadronic tensor and is proportional to the square of the 
hadronic part in Fig.~1(a) or the imaginary part of the forward 
scattering amplitude in Fig.~1(b).
The structure functions $F_1$ and $F_2$ contain all complex
hadronic interactions of the DIS cross section.

The structure functions are {\it locally} defined in DIS and can be 
directly extracted from measured DIS cross section via Eq.~(\ref{dis-x}).  
Because DIS cross section depends on both hard scattering scale $Q^2$ 
and soft momentum scale intrinsic to hadron wave function 
1/fm~$\sim\Lambda_{\rm QCD}$, structure functions $F_1$ and $F_2$ are
nonperturbative quantities.

%=======================================================================
\subsection{Parton distribution functions}

Parton distributions (or parton distribution 
functions), $\phi_{f/h}(x,\mu^2)$, are defined as matrix 
elements of a pair of parton fields and are often interpreted as the 
probability densities for finding a parton of flavor $f$ within a hadron 
$h$ of momentum $p$, with its momentum fraction between $x$ and $x+dx$ 
and it virtuality less than $\mu^2$ \cite{Collins:1981uw}.  
For example, a quark distribution is given by
\begin{equation}
\phi_{q/h}(x,\mu^2) 
= \int \frac{dy^-}{2\pi}\, {\rm e}^{ixp^+ y^-}
\langle h(p)| \bar{\psi}_q(0)\, \frac{\gamma^+}{2}
{{\cal P}{\rm e}^{-ig\int_0^{y^-}dw^-A^+(w^-)}}
\psi_q(y^-)|h(p)\rangle\ .
\end{equation}
Corresponding Feynman diagrams in momentum space are given by the 
type of cut-vertex diagrams in Fig.~2 \cite{Mueller:sg}.

Parton distributions are {\em universally} defined and in principle, 
in dependent of any specific physical process.  Because of the 
hadron state, like the structure functions, parton distributions are 
nonperturbative quantities.  However, unlike the structure functions, 
parton distributions are not {\it direct} physical observables. 

\begin{figure}[ht]
\v{-0.8cm}

\begin{minipage}[c]{16cm}
\centerline{\includegraphics[width=14.5cm]{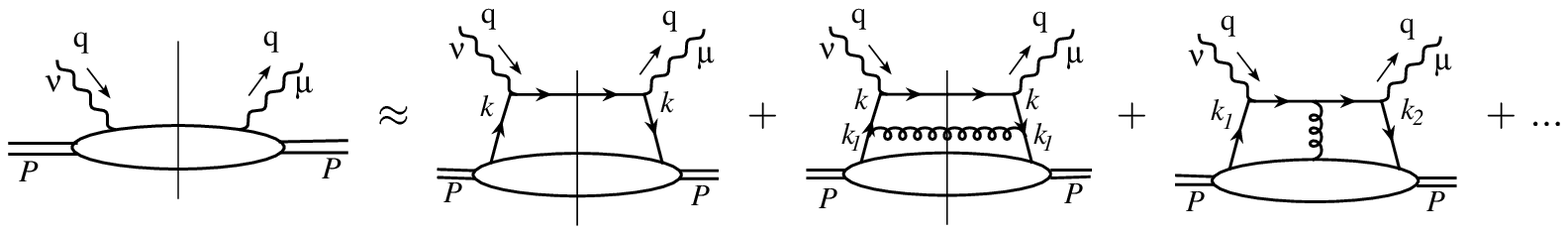}}
\centerline{Figure 3}
\end{minipage}
\end{figure}

\v{-1.3cm}
%=======================================================================
\subsection{Factorization}

Structure functions and parton distributions are related to each 
other via perturbative QCD factorization \cite{Curci:1980uw}.  

In QCD perturbation theory, the forward scattering amplitude in 
Fig.~1(b) can be represented in terms of an expansion of 
Feynman diagrams 
illustrated in Fig.~3. However, because of the soft momentum scale in
the hadron state, Feynman diagrams in Fig.~3 are not entirely 
calculable in perturbation theory.  For example, the integration of 
loop momentum $k$ for the leading order (LO) 
diagram can be expressed as

\v{-0.8cm}
\begin{figure}[ht]
\includegraphics[width=15cm]{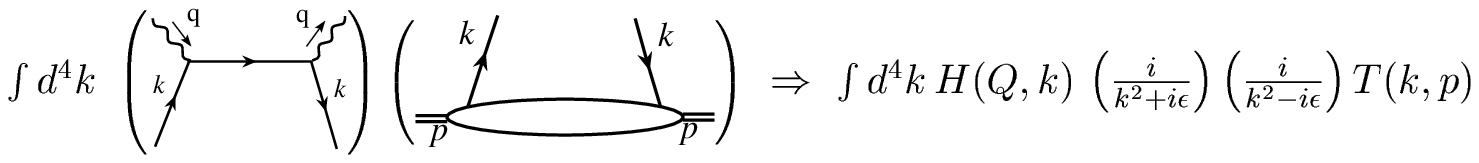}
\v{-0.8cm}

\end{figure}

\noindent
which is completely dominated by the phase space near 
$k^2\sim (1/\mbox{fm})^2 \sim \Lambda_{\rm QCD}^2$ 
because of the perturbative pinch 
singularity at $k^2\sim 0$.  Therefore, when $k^2\ll Q^2$, 
the dominant contribution
to the DIS cross section comes from the phase space where the 
active quark is ``long-lived'' relative to the time 
scale of hard collision, $t_c\sim 1/Q$.  It is such a ``long-lived'' 
parton state that separates long- from short-distrance effects
in the cross section. 

\begin{figure}[ht]
\v{-0.75cm}

\begin{minipage}[c]{16cm}
\centerline{\includegraphics[width=11.5cm]{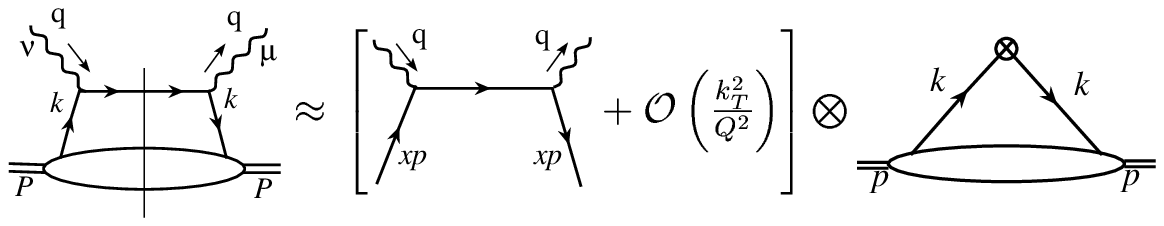}}
\centerline{Figure 4}
\end{minipage}
\v{-0.8cm}

\end{figure}

In terms of a power expansion of $\langle k^2\rangle/Q^2$, the LO
forward scattering amplitude in Fig.~3 can be approximated as 
illustrated in Fig.~4, and corresponding contribution to structure 
function $F_2$ can be expressed as
\begin{equation}
F_2(x_B,Q^2) =
x_B \sum_q e_q^2\ \phi_{q/h}(x_B)
+ {\cal O}\h{-0.02in}\left(\frac{\Lambda_{\rm QCD}^2}{Q^2}\right)\ .
\label{f2-born}
\end{equation}
If we neglect {\it all} high order diagrams and power corrections 
in $1/Q^2$, $F_2$ in Eq.~(\ref{f2-born}) has the Bjorken scaling in 
$x_B$ and is the same as the prediction of the parton model.

\begin{figure}[ht]
%\v{-0.7cm}

\begin{minipage}[c]{16cm}
\centerline{\includegraphics[width=15cm]{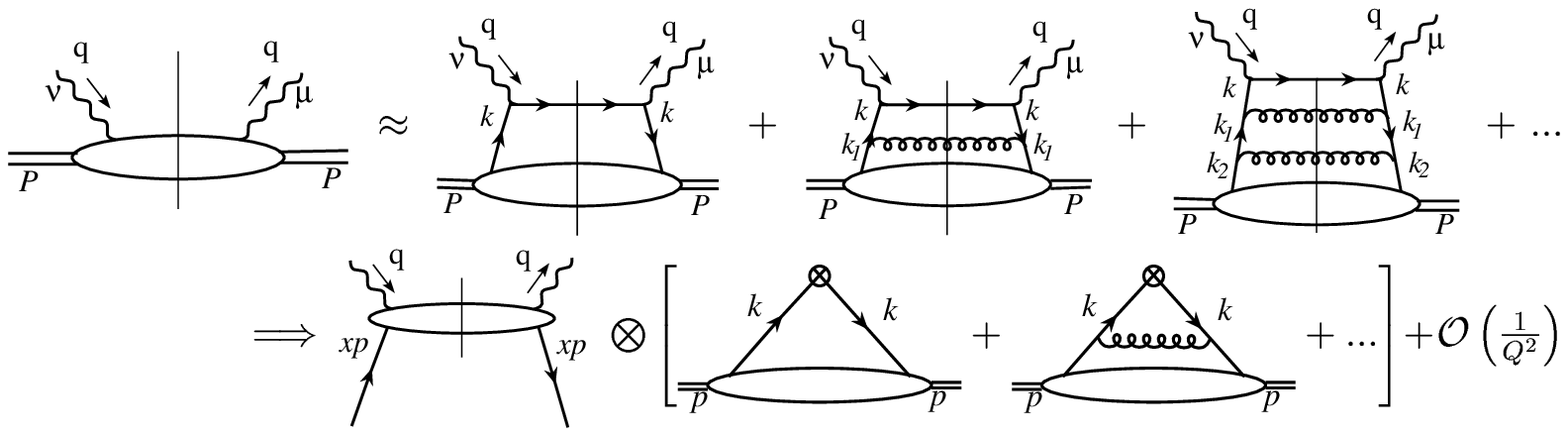}}

\vspace{0.08in}
\centerline{Figure 5}
\end{minipage}
\v{-0.8cm}

\end{figure}

However, QCD is much richer in dynamics than Feynman's parton model, 
and the rest diagrams in Fig.~3 will all contribute to the measured 
structure functions.  Although there are infinite diagrams, 
leading contributions to $W_{\mu\nu}$ in a physical gauge come from the 
diagrams with a ladder structure, as illustrated in Fig.~5, and partons'
loop momentum integrations $\int d^4k_i$ are dominated by the region 
of phase space near the perturbative pinch singularities, $k_i^2\sim 0$.
By neglecting the power corrections of $\langle k_i^2 \rangle/Q^2$, 
as illustrated in Fig.~5, 
the sum of all ladder diagrams in Fig.~5 can be factorized into a 
convolution of a short-distance part with all momentum scales of 
the order of $Q^2$ and a long-distance nonperturbative parton 
distribution \cite{Curci:1980uw},  
\begin{equation}
F_2(x_B,Q^2) = 
\sum_f \int_{x_B}^1 \frac{dx}{x}\,
       C_f\left(\frac{x_B}{x},\frac{Q^2}{\mu^2},\alpha_s(\mu)\right)\,
       \phi_{f/h}(x,\mu^2) 
+ {\cal O}\left(\frac{\Lambda_{\rm QCD}^2}{Q^2}\right)
\label{f2-qcd}
\end{equation}
where $\sum_f$ runs over all parton falvors, the coefficient 
functions $C_f$ are perturbatively calculable in a power series of 
$\alpha_s$, and the parton distributions $\phi_{f/h}$ represent 
the leading and universal part of the long-distance physics. 
All other nonperturbative contributions are powerly suppressed.

If the scattering is hard enough for neglecting all power corrections,
Eq.~(\ref{f2-qcd}) represents a fact that parton distributions, though 
theoretically defined, can be experimentally measured via the physical
quantities like the structure functions.  It also indicates
that a test of QCD in high energy collisions requires a good knowledge
of parton distributions.

%=======================================================================
\subsection{Global analyses of parton distributions}

Although parton distributions are nonperturbative, their
dependence on the factorization scale $\mu^2$ is predicted 
by pQCD in a form of DGLAP evolution equations 
\cite{DGLAP},
\begin{equation}
\mu^2 \frac{\partial}{\partial \mu^2} \phi_{i/h}(x,\mu^2)
= \sum_{j} 
\int_{x}^{1} \frac{dx'}{x'}\,
P_{i/j}\left(\frac{x}{x'},\alpha_s(\mu)\right)\,
\phi_{j/h}(x',\mu^2)
\label{DGLAP}
\end{equation}
with calculable splitting functions $P_{i/j}$ 
for $\mu^2 \gg \Lambda_{\rm QCD}^2$.  To solve for the
parton distributions, we need a set of input parton distributions at  
$\mu_0^2$, which can only be extracted 
from experimental data.

The global QCD analysis of parton distributions represents our effort
to find a best set of universal parton distributions from all 
existing data.  The analysis itself is an excellent 
test of QCD dynamics in the hard scattering, the factorization 
theorems, and the universality of parton distributions.  With the 
extracted parton distributions of a free nucleon, pQCD calculations with 
next-to-leading order (NLO) accuracy in $\alpha_s$ are consistent with 
thousands of data points from more than dozen physical observables
in hadonic collisions \cite{mrst,cteq6}.  
Since the factorization theorems determine the absolute normalization
for each observable, there is no need for any artificial $K$-factor.

%In the rest of this talk, I review the status of applying the 
%pQCD factorization theorems to hard probes in high energy collisions 
%involving nuclei.

%%%%%%%%%%%%%%%%%%%%%%%%%%%%%%%%%%%%%%%%%%%%%%%%%%%%%%%%%%%%%%%%%%%%%%%%
\section{Nuclear Structure Functions and Parton Distributions}

%=======================================================================
\subsection{Definitions}

The derivation of the DIS cross section in Eq.~(\ref{dis-x}) 
is independent of the details of the targets.  Nuclear structure 
functions, $F_i^A(x_B,Q^2)$, extracted from DIS data on a nuclear 
target of atomic weight $A$, are defined in the same way as that 
in Eq.~(\ref{f-wmn}) with the hadronic tensor of a nuclear state.
The Bjorken variable, $x_B \equiv \frac{Q^2}{2 p\cdot q}$
with an averaged nucleon momentum $p=\frac{P_A}{A}$,
and has a range from 1 to $A$.  

If the partons' typical virtuality in a nucleus, 
$\langle k^2 \rangle_A \ll Q^2$, and $\langle k_T \rangle_A \ll xp$, 
all derivations in last section for a free nucleon state can be
carried over for a nuclear state, and nuclear
structure function, $F_2^A$, shares the same factorized relation, 
\begin{equation}
F_2^{A}(x_B,Q^2) = 
\sum_f \int_{x_B}^1 \frac{dx}{x}\,
C_f\left(\frac{x_B}{x},\frac{Q^2}{\mu^2},\alpha_s(\mu)\right) 
\, \phi_{f/A}(x,\mu^2)
+ {\cal O}\h{-0.04in}\left(\frac{\langle k^2\rangle_A}{Q^2}\right)
\label{f2-qcd-A}
\end{equation}
where the short-distance coefficient functions $C_f$ should be the same 
as those for free nucleon and independent of $A$.
In Eq.~(\ref{f2-qcd-A}), 
nuclear parton distributions, $\phi_{f/A}(x,\mu^2)$, 
are matrix elements of nuclear states with the same operators of
nucleon parton distributions.  

If we can neglect the power corrections in Eq.~(\ref{f2-qcd-A}), 
i.e., pQCD factorization theorems hold, we can extract nuclear 
parton distributions and their $A$-dependence from the measured 
structure functions on nuclear targets.  Because of the universality 
of parton distributions, the extracted 
$A$-dependence of nuclear parton distributions should also be universal 
and represent the internal properties of a nuclear wave function.

\begin{figure}[ht]
\v{-0.7cm}

\begin{minipage}[c]{16cm}
\centerline{\includegraphics[width=11cm]{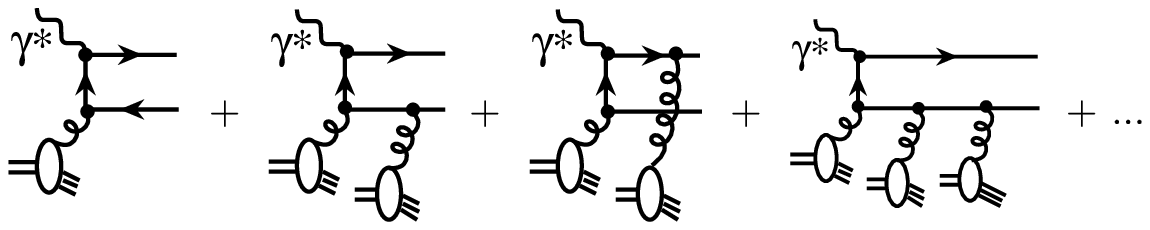}}

\vspace{0.07in}
\centerline{Figure 6}
\end{minipage}
\v{-1.0cm}

\end{figure}

%=======================================================================
\subsection{Nuclear shadowing}

Gluon distribution grows rapidly as momentum fraction $x$ decreases.  
When $x_B\ll 0.1$, a nucleus becomes a dense system of small-$x$ gluons 
and the large number of soft gluons from different
nucleons can all participate in the hard scattering, as illustrated 
in Fig.~6.  Such coherent multi-gluon interactions suppress
nuclear structure function $F_2^A(x_B,Q^2)$ in comparison to a sum
of free nucleon structure functions.  Such suppression is often referred
as nuclear shadowing, and it increases as $x_B$ decreases and/or 
$A$ increases.  

\begin{figure}[ht]
%\v{-0.8cm}

\begin{minipage}[c]{16cm}
\centerline{\includegraphics[width=15cm]{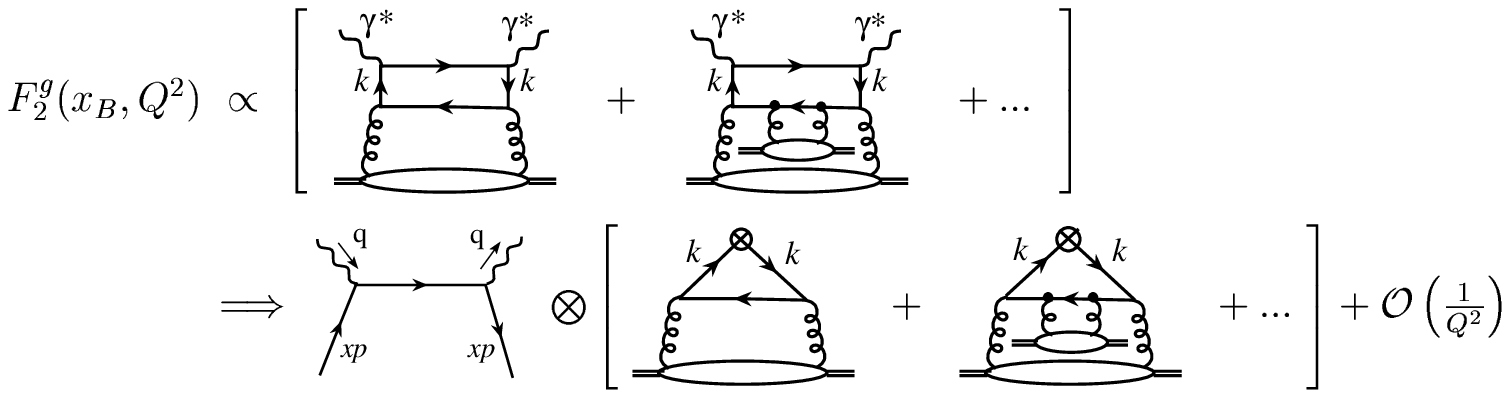}}

\vspace{0.07in}
\centerline{Figure 7}
\end{minipage}
\v{-0.8cm}

\end{figure}

In order to understand the phenomenon of nuclear shadowing in a language
close to pQCD factorization, let's consider a leading twist 
(or leading power) gluonic contribution to the structure function $F_2$, 
as illustrated in Fig.~7.  If the active quark's virtuality
in a large nucleus, $\langle k^2\rangle_A \ll Q^2$, the multiple
gluonic contributions are dominated by the phase space
where the quark state is ``long-lived'', which separates long- from
short-distrance effects, and pQCD factorization holds.
Unlike what shown in Fig.~5,  nuclear quark distribution in Fig.~7
gets contributions from multi-gluon interactions from different 
nucleons.  Since $\langle k^2\rangle_A \ll Q^2$, such multi-gluon 
interactions are remote from the short distance hard scattering, 
and thus, are internal to nuclear parton distributions.  

If the factorization scale $\mu^2$ is within perturbative region, 
multi-gluon interactions in Fig.~7 lead to calculable corrections to 
DGLAP equations.  The first corrections, in the small-$x$ limit, 
are known and lead to a set of modified evolution equations 
\cite{Mueller:wy}.  For gluon distribution in a nucleus, we have
\begin{equation}
\mu^2 \frac{\partial}{\partial \mu^2} xG_{A}(x,\mu^2)
= \frac{\alpha_sC_A}{\pi}
  \int_{x}^{1} \frac{dx'}{x'}\, x' G_A(x',\mu^2)
- \frac{\kappa}{R^2 \mu^2}
  \left(\frac{\alpha_sC_A}{\pi}\right)^2
  \int_{x}^{1} \frac{dx'}{x'}\, [x' G_A(x',\mu^2)]^2
\label{GLRMQ}
\end{equation}
where $\kappa$ is a known positive constant \cite{Mueller:wy}. 
The second term slows down the rapid growth of gluon distribution at
small-$x$ due to gluon recombination.  Although it is suppressed by
a power of $1/\mu^2$, the effect of this nonlinear term to gluon 
distribution does not really suppressed by $1/\mu^2$ because of 
a simple fact that $\int_{\mu_0^2}^{\mu^2} 
\frac{d\bar{\mu}^2}{(\bar{\mu}^2)^2} \rightarrow \frac{1}{\mu_0^2}$
as $\mu^2\rightarrow \infty$.  

According to pQCD factorization theorems, $A$-dependence of nuclear
parton distributions are universal, and can come from two sources:
(1) the nonlinear terms in the modified evolution equations and (2) 
input nuclear parton distributions at $\mu_0^2$, which have to be
extracted from experimental data.  It was shown \cite{Qiu:wh} that
the nonlinear evolution {\it in perturbative region} cannot be a main 
source to generate the observed nuclear shadowing, and 
parton distributions exhibit significant shadowing at the initial 
scale of evolution $\mu_0^2$.  However, for fixed small-$x$ the 
nonlinear evolution is sufficient to keep the shadowing going away
very slowly as one increases $\mu^2$ even up to values as large as 
100~GeV$^2$.  This is a solid prediction of perturbative QCD and 
has been confirmed experimentally.

On the other hand, $A$-dependence of nuclear structure functions
are not universal because of the power corrections in 
Eq.~(\ref{f2-qcd-A}). 

%=======================================================================
\subsection{Saturation -- breakdown of pQCD factorization}

PQCD factorization for nuclear structure functions in Eq.~(\ref{f2-qcd-A}) 
is only valid for a dilute nucleus.  If the direct power corrections in 
Eq.~(\ref{f2-qcd-A}) and/or the power corrections to the modified evolution
in Eq.~(\ref{GLRMQ}) become important it is a signal that a dense system
of partons is reached.  When the power 
corrections are comparable to the leading power contributions, the 
conventional pQCD factorization breaks down.  

Although Eq.~(\ref{GLRMQ}) is really valid only when the nonlinear
term is small compared to the usual evolution term.  Nevertheless,
Eq.~(\ref{GLRMQ}) is an interesting equation in that the nonlinear 
term stabilizes the growth of normal evolution and leads to a limiting
value for $xG_A(x,\mu^2)$ as $x\rightarrow 0$.  We can estimate
roughly where this saturation sets in by finding where the gluon 
distribution loses its $x$-dependence, and find 
$xG(x,Q_s^2)\propto R^2 Q_s^2 / \alpha_s(Q_s)$ at 
a saturation scale $Q_s^2$.  More recent developments in understanding 
saturation can be found in Ref.~\cite{AHM-qm2002}.  

\begin{figure}[ht]
%\v{-0.7cm}

\begin{minipage}[c]{16cm}
\centerline{\includegraphics[width=15cm]{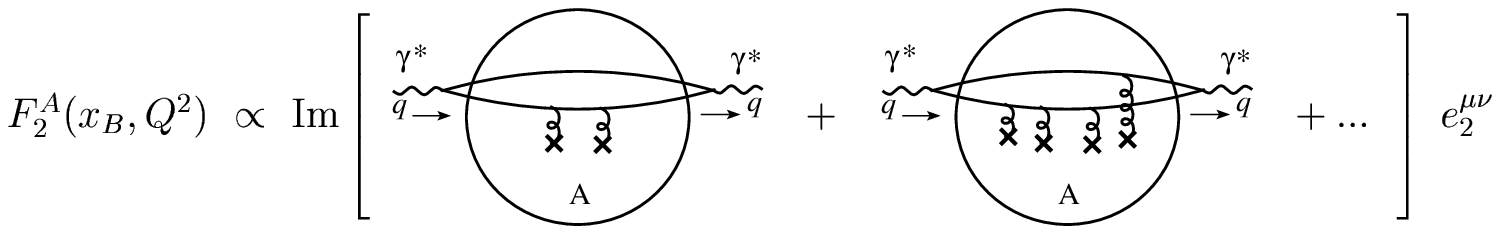}}

\vspace{0.06in}
\centerline{Figure 8}
\end{minipage}
\v{-0.8cm}

\end{figure}

When $x_B\rightarrow 0$, all order gluonic contributions to the 
structure functions, as illustrated in Fig.~6, are equally important.  
If the intrinsic parton virtuality in such a dense system, 
$\langle k^2\rangle_A \sim Q_s^2 \gg \Lambda_{\rm QCD}^2$, the active 
partons are ``short-lived'', and multiple gluonic contributions to
the structure functions in Fig.~6 or 7 might be calculated without 
introducing a ``sea'' quark distribution.   
The calculation can be carried out 
in the target rest frame as illustrated in Fig.~8, 
and the saturated nuclear structure function $F_2^A$ is given 
by \cite{AHM-qm2002}
\begin{equation}
F_2^A \propto
\int d^2b\, dz\,
\left|\psi_{\gamma^*\rightarrow q\bar{q}}(b,z,Q^2)\right|^2  
\sigma_{q\bar{q}-A}(b,z,Q_s^2)
\end{equation}
where $\psi_{\gamma^*\rightarrow q\bar{q}}$ is the wavefunction for
a virtual photon to go into a quark-antiquark pair of longitudinal
momentum fractions $z$ and $1-z$, $b$ is a Fourier conjugate of 
relative momentum between $q$ and $\bar{q}$, and 
$\sigma_{q\bar{q}-A}$ reprensents a hadronic cross section between
the $q\bar{q}$ pair and the nucleus.  
%More discussions can be found in Ref.~\cite{AHM-qm2002}.

In this saturation regime, the pQCD factorization formula in 
Eq.~(\ref{f2-qcd-A}) is not valid.  What calculated in this saturation
limit are nuclear structure functions 
including all power corrections, not the parton distributions,
which, by definition, are twist-2 matrix elements and universal.

%=======================================================================
\subsection{Factorization in heavy ion collisions}

For a hard probe of scale $Q$ in relativistic heavy ion collisions, 
there could be three types of multi-gluon 
interactions: (1) within individual ion, (2) between the hard scattering 
and one or both ions, and (3) between two ions, 
as illustrated in Fig.~9.  

\begin{figure}[ht]
\v{-0.6cm}

\begin{minipage}[c]{16cm}
\centerline{\includegraphics[width=14cm]{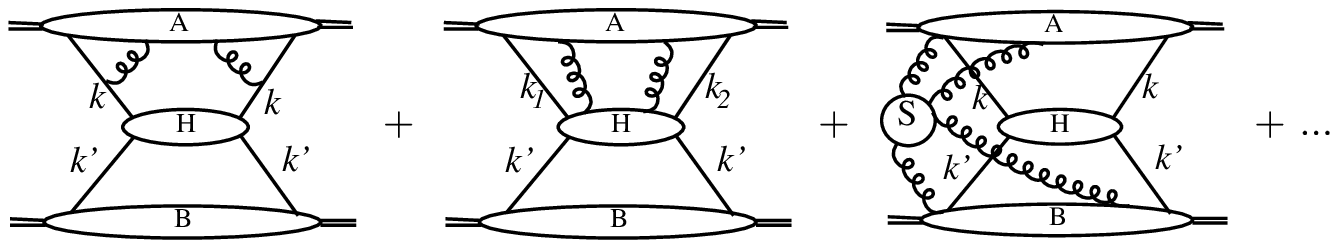}}

\vspace{0.08in}
\centerline{Figure 9}
\end{minipage}
\v{-0.7cm}

\end{figure}

If the parton virtuality $k^2\ll Q^2$, the first type soft gluon 
interactions are remote from the hard probe and internal to 
individual ion, and therefore, should have been included in
nuclear parton distributions and do not interfere with pQCD 
factorization.  The second type of multi-gluon
interactions with physical polarizations are suppressed by powers of 
$\langle k^2\rangle/Q^2$ because they link effects at two different 
scales.  If soft gluon interactions with the hard part involve only 
one ion, a genralized factorization for leading power corrections 
should be valid for calculating this type of interactions 
\cite{Qiu:xy}.

The third type of multi-gluon interactions is the most
difficult one to remove.  Such soft gluon interactions exchange 
informations between two ions, and therefore, 
have a potential to alter the parton distributions
before the hard scattering takes place.  Without the universality of 
parton distributions, pQCD calculations do not have real predictive 
power.  Due to the gauge invariance of QCD and unitarity, 
contributions of the third type soft gluon interactions to a physical 
observable are suppressed by a power of ($1/Q^4$)
\cite{Qiu:xy}.

Therefore, pQCD factorization formalism for hadronic collisions 
can apply to hard probes in heavy ion collisions so long as the 
power corrections are relatively small,
\begin{equation}
\frac{d\sigma_{AB}}{dQ^2} 
= \sum_{a,b}
\phi_{a/A}(x_a,\mu^2) \otimes
\phi_{b/B}(x_b,\mu^2) \otimes
 H_{ab}\left(x_a x_b S,\frac{Q^2}{\mu^2}, \alpha_s(\mu)\right) 
+ {\cal O}
\left(\frac{\langle k^2\rangle}{Q^2}\right)
\label{AB-fac}
\end{equation}
where $\otimes$ repesents convolution over parton momentum fractions 
$x_a$ and $x_b$.  One 
can also include nuclear thickness functions into Eq.~(\ref{AB-fac}) to 
take care of the size effect of the ions.  At this leading power level, 
$A$-dependence of an observable is completely determined 
by nuclear parton distributions.

When the hard scale $Q\ll S$, momentum fractions of the active partons 
can be very small.  Soft gluon interactions between two ions, in addition
to the interactions within individual ion, can lead to a much larger
$Q_s^2$ in heavy ion collisions than that in DIS.  When $Q$ is not too 
much larger than $Q_s$, pQCD factorization fails, and normal concept of
parton distributions cannot be applied to such a dense system.

If the power corrections are significant, but not too large,  
pQCD calculations might be systematically carried out
at the first power corrections.  The predictions are often expressed in
terms of multi-parton correlation functions \cite{Qiu:2001hj}.

%%%%%%%%%%%%%%%%%%%%%%%%%%%%%%%%%%%%%%%%%%%%%%%%%%%%%%%%%%%%%%%%%%%%%%%%
\section{Global Analysis of Nuclear Parton Distribution Functions}

%Although there are several attempts to describe the nuclear modifications
%to parton distributions.  In this talk, I concentrate on global analysis
%of nuclear parton distributions.

%=======================================================================
\subsection{Methodology}

Predictive power of pQCD calculations of hard probes in heavy ion 
collisions relies on a set of universal nuclear parton distributions.
Therefore, a reliable set of nuclear parton distributions
is crucial for understanding RHIC physics.  Any new physics and phenomena 
should represent the observations beyond what predicted by pQCD 
factorization.

PQCD predicts the scale dependence of parton distributions in terms
of evolution equations.  Before performing a global QCD analysis of 
nuclear parton distributions, we are required to choose: (1) evolution 
equations, (2) parameterizations of input nuclear parton distributions at 
a scale $\mu_0^2$, and (3) data sets with rich nuclear information.
Then, we need to (1) evolve the input distributions to any other values 
of $\mu^2$, (2) calculate the theoretical 
predictions by using the evolved distributions, and (3) 
compare the predictions with real data and calculate the $\chi^2$,
The best set of nuclear parton distributions should give a minimum 
$\chi^2$ or a best fit to all existing data.  

At the leading twist level, $A$-dependence of all physical observables
are completely determined by the nuclear dependence of the parton
distributions.  Deriving a reliable $A$-dependence of parton 
distributions is a most crucial part of the global analysis.  
There are two possible approaches to derive the $A$-dependence: 
(A) DGLAP evolution with all $A$-dependence included in
the input distributions at $\mu_0^2$ and (B) modified evolution 
equations of the type in Eq.~(\ref{GLRMQ}) with nonperturbative 
$A$-dependence included in the input distributions at $\mu_0^2$ and 
perturbative $A$-dependence generated from the evolution.

%=======================================================================
\subsection{Exsiting work}

Two groups have been doing global analysis of nuclear parton 
distributions.  Eskola, Kolhinen, Ruuskanen, and Salgado (EKRS) 
produced EKS98 package of nuclear parton distributions \cite{EKS98}.
Hirai, Kumano, and Miyama (HKM) derived several sets of 
nuclear parton distributions from extensive DIS data \cite{HKM}.

%=======================================================================
% similarities and differences of two groups

Both groups used LO DGLAP evolution equations and adopted 
the approach (A) for including nuclear dependence.
They define the input parton distributions as
\begin{equation}
\phi_{f/A}(x,\mu_0^2) = R_f^A(x,\mu_0^2)\, 
 \phi_{f/N}(x,\mu_0^2)
\end{equation}
with $\phi_{f/N}(x,\mu_0^2)$ are known free nucleon parton 
distributions.  EKRS used both CTEQ-LO and GRV-LO free nucleon 
parton distributions for producing EKS98, 
while HKM used MRST-LO parton distribution for its nuclear parton 
distributions.  Two groups used different parameterizations for 
$R_f^A(x,\mu_0^2)$.  The goal of the global analysis is to extract the 
ratio $R_f^A(x,\mu_0^2)$ that represents a best fit to all data used 
in the analysis.

For data selection, two groups used different data sets to produce 
their published nuclear parton distributions \cite{EKS98,HKM}.
EKRS used both DIS and Drell-Yan data on nuclear targets.  In 
particular, Drell-Yan data from Fermilab E772 experiment help to
fix relative nuclear effects in valence and sea and make 
the ratio for sea quarks, $R_S(x,\mu_0^2) < 1$ at medium $x$.  
In addition, EKRS used NMC data on $Q^2$-dependence of 
$F_2^{\rm Sn}/F_2^{\rm C}$ that provide constraints 
on $R_g^A(x,\mu_0^2)$. On the other hand, 
HKM used only DIS data.  

Detailed comparison between data and theoretical calculations using 
EKRS nuclear parton distributions and other available parameterizations 
can be found in Ref.~\cite{EKRS-data}. 
Similar comparison for HKM distributions can be found in 
Refs.~\cite{HKM,EKRS-data}.  
Both published EKRS and HKM nuclear parton distributions can fit 
their selected data sets very well. 
I summarize a few key differences here:
(1) there are large differences in input distributions at 
$\mu_0^2=2.25$~GeV$^2$; 
(2) the default set of nuclear parton distributions
in HIJING seems to give a too strong $A$-dependence; and (3) 
HKM distributions appear to predict a different $Q^2$-dependence 
for NMC data on the ratio $F_2^{\rm Sn}/F_2^{\rm C}$ and a 
different shape in $x_2$ dependence for E772 Drell-Yan data.  

Notice that both of these data
were not included in the HKM's original global analysis. 
Since DIS data are only sensitive to the sum of quark and antiquark 
distributions, it should not be surpprised that the separation of 
sea and valence quarks in HKM distributions is not 
very well constrained.  Because LO structure function $F_2$ does not 
have an explicit dependence on gluon distribution, gluons  
could not be well constrained by DIS data alone.   
These two ambiguities are reduced in EKRS approach because of 
the usage of Drell-Yan data and NMC data on $Q^2$-dependence of 
$F_2^{\rm Sn}/F_2^{\rm C}$.

%%%%%%%%%%%%%%%%%%%%%%%%%%%%%%%%%%%%%%%%%%%%%%%%%%%%%%%%%%%%%%%%%%%%%%%%
\section{Summary and Outlook}

Much of the predictive power of pQCD calculations relies on the 
factorization and universality of parton distributions.  
PQCD calculations with next-to-leading order accuracy in $\alpha_s$ 
have been consistent with almost all existing data 
from hadronic collisions without a need for an artifical $K$-factor.

PQCD factorization theorems can be applied to hard scattering involving 
nuclei, if the partonic system covered by the interaction volume is
a dilute system, in which partons are ``free'' or ``long-lived'' 
at the time scale of hard collision.  If the leading twist pQCD
factorization theorems hold, $A$-dependence of all observables
are completely determined by the universal $A$-dependence of 
nuclear parton distributions.  Any significant deviation from the 
predicted $A$-dependence is a signal of breakdown of the conventional
pQCD factorization and new physical phenomena.  Therefore, 
precise $A$-dependence of nuclear parton distributions are
very important for understanding hard scattering signals in heavy
ion collisions at RHIC and the LHC energies.

Nuclear dependence of the input distributions is a main 
source of the $A$-dependence of nuclear parton distributions and has
to be extracted from better data. 
At high energy or small-$x$, perturbative $A$-dependence via 
the modified evolution equations might be important too.
Two sets of nuclear parton distributions from LO global QCD analysis 
are available.  In order to reach the same level of precision tests 
that achieved in collisions with free hadrons, we need NLO nuclear 
parton distributions with better precisions on the $A$-dependence.

%%%%%%%%%%%%%%%%%%%%%%%%%%%%%%%%%%%%%%%%%%%%%%%%%%%%%%%%%%%%%%%%%%%%%%%%

%%%%%%%%%%%%%%%%%%%%%%%%%%%%%%%%%%%%%%%%%%%%%%%%%%%%%%%%%%%%%%%%%%%%%%%%
\end{document}